\documentclass[reprint,superscriptaddress, aps,prb,]{revtex4-1}

\usepackage{graphicx}
\usepackage{dcolumn}
\usepackage{bm}
\usepackage{hyperref}

\begin{document}

\title{Relation between cuprate superconductivity and magnetism:\\ A Raman study of (CaLa)(BaLa)$_{2}$Cu$_{3}$O$_{y}$}

\author{Dirk Wulferding}
\affiliation{Center for Artificial Low Dimensional Electronic Systems, Institute for Basic Science, 77 Cheongam-Ro, Nam-Gu, Pohang 790-784, Korea}
\affiliation{Department of Physics, Pohang University of Science and Technology, Pohang 790-784, Korea}
\affiliation{Institute for Condensed Matter Physics, Technical University of Braunschweig, D-38106 Braunschweig, Germany}

\author{Meni Shay}
\affiliation{Department of Physics, Technion - Israel Institute of Technology, Haifa 32000, Israel}
\affiliation{Department of Physics and Optical Engineering, Ort Braude College, 21982 Karmiel, Israel}

\author{Gil Drachuck}
\affiliation{Department of Physics, Technion - Israel Institute of Technology, Haifa 32000, Israel}

\author{Rinat Ofer}
\affiliation{Department of Physics, Technion - Israel Institute of Technology, Haifa 32000, Israel}

\author{Galina Bazalitsky}
\affiliation{Department of Physics, Technion - Israel Institute of Technology, Haifa 32000, Israel}

\author{Zaher Salman}
\affiliation{Laboratory for Muon Spectroscopy, Paul Scherrer Institute, 5232 Villigen PSI, Switzerland}

\author{Peter Lemmens}
\affiliation{Institute for Condensed Matter Physics, Technical University of Braunschweig, D-38106 Braunschweig, Germany}

\author{Amit Keren}
\affiliation{Department of Physics, Technion - Israel Institute of Technology, Haifa 32000, Israel}

\date{\today}

\begin{abstract}

We present an investigation of charge-compensated antiferromagnetic (Ca$_{x}$La$_{1-x}$)(Ba$_{1.75-x}$La$_{0.25+x}$)Cu$_{3}$O$_{y}$ single crystals using Raman scattering as well as muon spin rotation. In this system the parameter $x$ controls the Cu-O-Cu superexchange interaction via bond distances and buckling angles. The oxygen content $y$ controls the charge doping. In the absence of doping the two-magnon peak position is directly proportional to the superexchange strength $J$. We find that both $x$ and $y$ affect the peak position considerably. The N\'{e}el temperature determined from muon spin rotation on the same samples independently confirms the strong dependence of the magnetic interaction on $x$ and $y$. We find a considerable increase in the maximum superconducting transition temperature $T_{c}^{max}$ with $J$. This is strong evidence of the importance of orbital overlap to superconductivity in this family of cuprates.

\begin{description}
\item[PACS numbers] {74.20.-z, 74.25.Ha, 74.72.Cj}
\end{description}
\end{abstract}

\maketitle

\section{Introduction}

The complexity of high-temperature superconductors (HTSC), with several parameters controlling their properties, hinders progress in uncovering their superconducting mechanism. It is extremely difficult to isolate and control only one parameter at a time. To overcome the experimental challenge in the study of HTSC, we prepare and investigate single crystalline samples of the (Ca$_{x}$La$_{1-x}$)(Ba$_{1.75-x}$La$_{0.25+x}$)Cu$_{3}$O$_{y}$ (CLBLCO) system, with $x=0.1$, $0.2$, and $0.4$. This system is isostructural to YBa$_2$Cu$_3$O$_{7-y}$ (YBCO) and supplies us with two tuning parameters: the amount of oxygen $y$ and the ratio of calcium to barium set by $x$. Each $x$ defines a superconducting family. Calcium and barium have the same valence, and their total amount in the chemical formula is constant. Therefore, $x$ does not serve as a formal dopant. However, $x$ tunes the Cu-O-Cu bond distances and buckling angles, leading to variations in the overlap of orbitals.~\cite{Rinat08} This property of CLBLCO opens a window to investigate the relation between $T_{c}$ and the physical parameters that are determined by orbital overlap.

\begin{figure}[tbp]
\label{figure1}
\centering
\includegraphics[width=8cm]{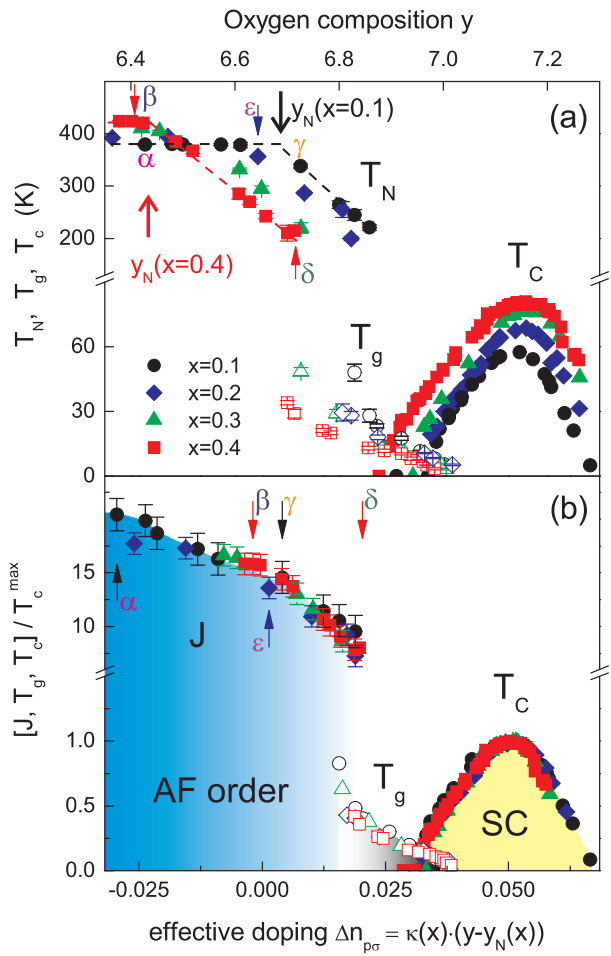}
\caption{(Color online) (a) Phase diagram of CLBLCO for the four families $x$=0.1, 0.2, 0.3, and 0.4. $T_{N}$, $T_{g}$ (open symbols) and $T_{c}$ are plotted as a function of oxygen composition $y$. The long arrows point to the $y$ values where $T_{N}$ starts to drop, which defines $y_{N}$. The samples $\alpha =(0.1,6.4)$, $\beta =(0.4,6.4)$, $\gamma =(0.1,6.75)$, $\delta =(0.4,6.73)$, and $\varepsilon =(0.2,6.64)$ are marked by their $(x,y)$ values. (b) $J$ and the critical temperatures normalized by $T_{c}^{max}$ as a function of doping variation $\Delta n_{p\sigma }$. $\alpha $ -- $\varepsilon $ indicate the positions of the five samples in the phase diagram.}
\end{figure}

Phase diagrams of CLBLCO with different $x$ are shown in Fig. 1(a). They summarize muon spin rotation ($\mu$SR)~\cite{Kanigel02,Rinat06} and transport~\cite{Yaki99} experiments performed on powder samples. It can be seen that the transition temperatures to the antiferromagnetic, spin-glass, and superconducting phases ($T_{N}$, $T_{g}$, and $T_{c}$, respectively) are strongly $x$ dependent. In particular, the superconducting transition temperature of the optimally doped samples $T_{c}^{max}$ increases by more than 30\% from the $x=0.1$ family to the $x=0.4$ family. However, high-resolution powder x-ray diffraction~\cite{Xray} and NMR experiments~\cite{KerenNJP09} indicate that $x=0.1$ samples are more ordered than $x=0.4$ ones. Therefore, disorder can not be responsible for the reduction of $T_{c}^{max}$ as $x$ decreases; there has to be a more fundamental reason for these variations of $T_{c}^{max}$.

Figure 1(b) depicts a re-scaled phase diagram for CLBLCO. We obtain this unified diagram by replacing $T_{N}$ with the superexchange strength $J$, normalizing the temperature scales by $T_{c}^{max}$,~\cite{Rinat06} and transforming the oxygen composition parameter $y$ according to $\Delta n_{p\sigma}=\kappa (x) (y-y_{N})$.~\cite{Eran10} The scaling parameter $\kappa (x)$ translates to the doping efficiency with which holes are introduced in the CuO$_{2}$ planes and was determined via measurements of the $^{17}$O nuclear quadrupole resonance parameter $\nu _{Q}$.~\cite{Eran10} $y_{N}$ is the oxygen composition where $T_{N}$ starts to decrease, as shown in Fig. 1(b), and $\Delta n_{p\sigma}$ stands for the effective hole variation on the oxygen orbital. The unified phase diagram suggests a similar origin for the magnetic and superconducting phase transitions. The superexchange $J$ is determined by orbital overlap and is closely related to the hopping rate $t$. While $J$ is well-defined for only the parent compounds, hopping exists even in the doped system. A close relation between $T_{c}$ and $J$ would therefore imply that superconductivity in this family of cuprates is dominantly driven by kinetic energy.~\cite{molegraaf}

However, the values for $J$ were determined by $\mu$SR. This method requires measurements over a wide range of temperatures in which $J$ could change due to lattice expansion. Furthermore, the theory that extracts $J$ from the data is rather involved and can only fit the data to a certain extent.~\cite{Rinat06} The work presented here overcomes these challenges by using two-magnon Raman scattering experiments. This technique determines $J$ directly by measuring the shift in photon energy $E_{max}$ between incoming and outgoing light due to a spin-flip process of two adjacent spins. Our measurements of the Raman shift as a function of both $x$ and $y$ reveal a strong correlation between $T_{c}^{max}$ and $J$ in CLBLCO.

\section{Experimental details}

Several single crystals of (Ca$_{x}$La$_{1-x}$)(Ba$_{1.75-x}$La$_{0.25+x}$)Cu$_{3}$O$_{y}$ with varying $y$ for both the $x=0.1$ and $x=0.4$ families, as well as one single crystal of the $x=0.2$ family, were successfully grown in a floating zone furnace.~\cite{Crystal} In order to control the amount of oxygen in the samples we followed the procedure known for powder samples of this material.~\cite{Yaki99} The oxygen composition of the samples has been determined by iodometric titration.~\cite{Crystal}

For the Raman measurements the crystals were cleaved to obtain a shiny, flat, virgin surface. The cleaving produces facets which are perpendicular to the $c$ axis. Therefore, the light scatters with its polarization within the $ab$ plane for all measurements. Polarized Raman spectra were obtained using a Jobin-Yvon micro-Raman spectrometer (LabRAM HR) in backscattering geometry with a $\lambda=532$ nm solid-state Nd:YAG laser, a 50$\times$ magnification objective, and a diffraction grating with 1800 grooves/mm. The detection was done with a nitrogen-cooled CCD (Horiba Spectrum One). To prevent damage to the sample from overheating the laser power at the sample was kept below 1 mW.

\section{Experimental Results and Discussion}

The CLBLCO system crystallizes in a simple tetragonal structure (space group $P4/mmm$). Five Raman-active phonon modes are expected from backscattering within the $ab$ plane. Four of these phonon modes are of $A_{1g}$ symmetry, where the scattered light has the same polarization as the incident light.~\cite{Crystal} One phonon mode has a $B_{1g}$ symmetry, where the polarization of the scattered light $\vec{e_{s}}$ is perpendicular to the incident light polarization $\vec{e_{i}}$ and the phonon intensity reaches a maximum when both $\vec{e_{i}}$ and $\vec{e_{s}}$ are at $45^{\circ}$ from the $a$ and $b$ axes. This intensity dependence allows for an easy orientation of the crystallographic $ab$ plane.

\begin{figure}[tbp]
\label{figure2}
\centering
\includegraphics[width=8cm]{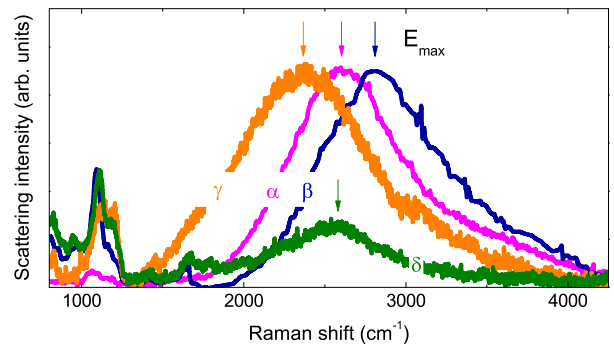}
\caption{(Color online) The two-magnon mode for samples $\alpha$ -- $\delta$, measured at $T=20$ K.}
\end{figure}

In Fig. 2 we depict raw two-magnon Raman scattering data for four samples marked $\alpha$, $\beta$, $\gamma$, $\delta$. The position of these samples in the phase diagram is indicated in Figs. 1(a) and 1(b). Samples $\alpha$ and $\gamma$ are part of the $x=0.1$ family, while samples $\beta$ and $\delta$ belong to the $x=0.4$ family. In the unified phase diagram [Fig. 1(b)], it can be seen that samples $\beta$ and $\gamma$ are comparable with respect to doping at $\Delta n_{p\sigma}\approx 0$, while sample $\alpha$ is in the highly underdoped region and sample $\delta$ is on the verge of the spin-glass phase.

The data in Fig. 2 are taken at $T=20$ K, deep in the magnetically ordered phase. Peaks with energies below 1500 cm$^{-1}$ are due to phonon and multiphonon scattering. The arrows in Fig. 2 indicate the peaks' respective maxima. A clear shift in the two-magnon peak energy is observed for the different samples. For both families the peak shifts to lower energies as $y$ increases, in accordance with a previous Raman study.~\cite{Sugai} As $y$ approaches the spin-glass phase, the two-magnon Raman signal dramatically decreases in intensity and vanishes in the spin-glass phase. At equal $y$, the $x=0.4$ sample has a higher two-magnon mode energy than the $x=0.1$ sample. The principle observation is that samples $\beta$ and $\gamma$, with the same effective doping $\Delta n_{p\sigma}$ but varying $x$, exhibit a huge difference in the two-magnon energy, with $\beta$ having the largest energy.

\begin{figure}[tbp]
\label{figure3}
\centering
\includegraphics[width=8cm]{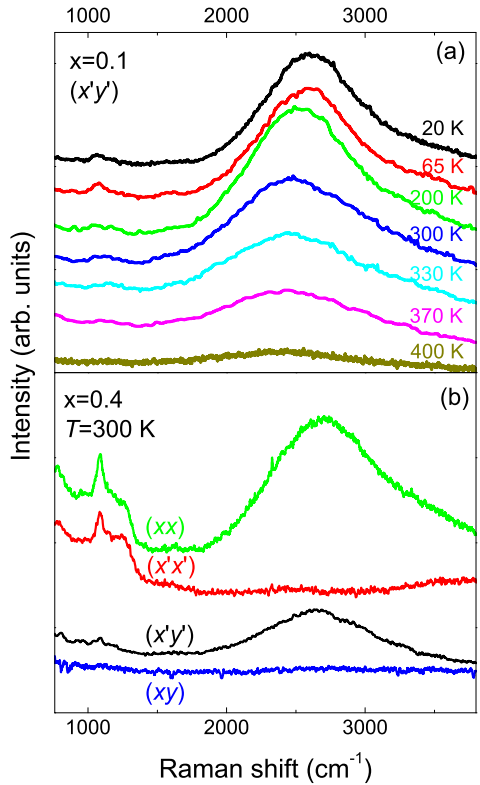}
\caption{(Color online) (a) Temperature dependence of the two-magnon mode in ($x'y'$) polarization for the $x=0.1$ sample. (b) Polarization dependence of the two-magnon mode measured at $T=300$ K.}
\end{figure}

In order to verify that the broad peak around 2500 cm$^{-1}$ is indeed related to a two-magnon scattering process we measure its intensity as a function of temperature. The results for $x=0.1$, $y=6.4$ are shown in Fig. 3(a). It is evident that the intensity of the peak drops above $T_{N}=376$ K. However, some remaining intensity of the two-magnon scattering is observed even at 400 K due to its local scattering nature. Moreover, as the temperature increases, the peak shifts to lower energies, revealing the temperature dependence of $E_{max}$. At $T\simeq T_{N}$ there is a high probability that neighbors of the spin-flipping pair are excited. This will reduce the energy cost in the flipping process.~\cite{TDepRaman}

Figure 3(b) shows the Raman spectra at four different polarizations. Here, we use the following notations: ($xx$) $\mathrel{\hat=}\vec{e_{i}}$ and $\vec{e_{s}}\parallel $ $a$ axis, ($x^{\prime}x^{\prime}$) $\mathrel{\hat=}\vec{e_{i}}$ and $\vec{e_{s}}\angle 45^{\circ}$ $a$ axis, ($x^{\prime}y^{\prime}$) $\mathrel{\hat=}\vec{e_{i}}\perp \vec{e_{s}}\angle 45^{\circ }$ $b$ axis, and ($xy$) $\mathrel{\hat=}\vec{e_{i}}\perp \vec{e_{s}}\parallel $ $b$ axis. It is found that the broad peak appears only in the $B_{1g}$ configuration, i.e., in ($xx$) and ($x^{\prime}y^{\prime}$) polarization, as expected from two-magnon Raman scattering in a square lattice.~\cite{Fleury,Freitas}

In Fig. 4(a) we plot the low-energy part of the Raman spectra for samples $\alpha$ and $\beta$, focusing on the $B_{1g}$ phonon mode. The atomic displacement for this mode corresponds to an out-of-plane motion of oxygen ions in the CuO$_{2}$ plane at $\approx$300 cm$^{-1}$. This motion is sketched in the inset of Fig. 4(a). We note two important observations regarding the phonons: ($i$) the $x=0.1$ sample has a slightly higher phonon frequency, and ($ii$) the family with $x=0.1$ has a smaller line-width than $x=0.4$ or $x=0.2$, which is also indicated. Both trends are independent of doping. In particular, the $y$ dependence of the line-width can be seen in Fig. 4(b).

Observation ($i$) supports the notion that the $\approx$300 cm$^{-1}$ phonon does not play a dominating role in the superconducting mechanism for CLBLCO since a higher phonon frequency should lead to a higher $T_{c}$. The respective line-width is a measure of the crystal quality and the electron-phonon coupling strength. It is inversely proportional to the phonon life time. Thus, observation ($ii$) suggests that as $x$ decreases, the coherence does not change or even increases. This is in agreement with the x-ray diffraction and NMR measurements of CLBLCO and reinforces the conclusion that the crystal quality is not responsible for the variation in $T_{c}^{max}$.

\begin{figure}[tbp]
\label{figure4}
\centering
\includegraphics[width=8cm]{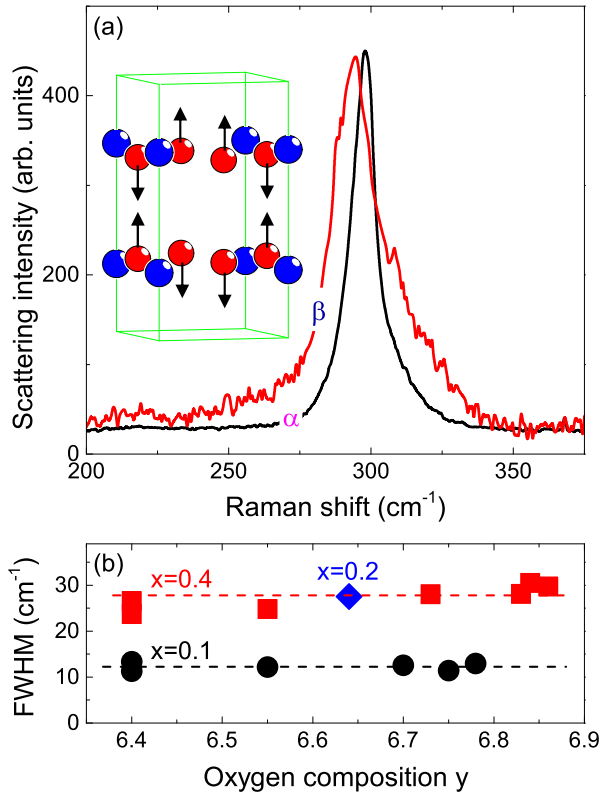}
\caption{(Color online) (a) Comparison of the $B_{1g}$ phonon mode of samples $\alpha$ and $\beta$. The atomic displacement pattern for this phonon mode is indicated in the inset; the red and blue spheres correspond to oxygen and copper, respectively. (b) Full width at half maximum (FWHM) of the $B_{1g}$ phonon mode at 300 cm$^{-1}$ for both $x=0.1$ and $0.4$ families as a function of oxygen composition $y$. The $x=0.2$ sample is indicated as well. The lines are guides to the eyes.}
\end{figure}

\begin{figure}[tbp]
\label{figure5}
\centering
\includegraphics[width=8cm]{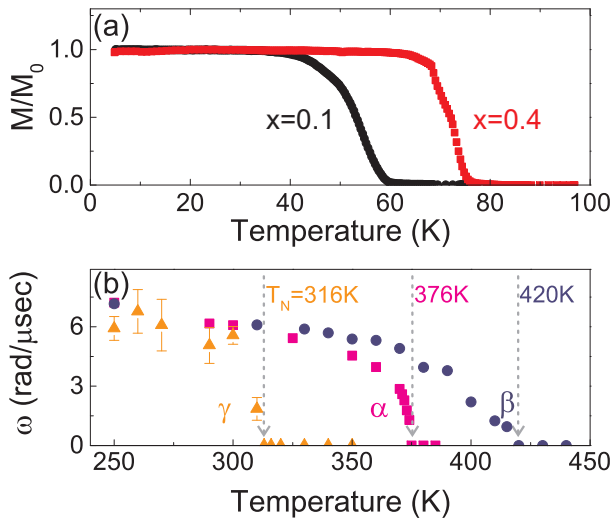}
\caption{(Color online) (a) Magnetization as function of temperature for optimally doped samples of $x=0.1$ and $x=0.4$. (b) N\'{e}el temperature determined for samples $\alpha$, $\beta$, $\gamma$ as indicated in the phase diagram using $\mu$SR.}
\end{figure}

Two of the samples that were used in the Raman measurements were subsequently oxygenated close to optimum doping at $y=7.12$. Their magnetization curves, obtained from a superconducting quantum interference device (SQUID) magnetometer, are shown in Fig. 5(a). A clear transition from the normal state to the superconducting state can be seen, with $T_{c}$ of the $x=0.4$ sample being much higher than that of the $x=0.1$ sample. As single crystals that are oxygenated to the exact optimum doping are difficult to achieve, we will use the values for $T_{c}^{max}$ obtained from the phase diagram in Fig. 1(a) in the following.

Since the oxygenation of crystals is slow and Raman scattering is surface sensitive, it is also important to check the antiferromagnetic part of the phase diagram. Thus, we determine $T_{N}$ for the samples that participated in the Raman experiments using zero-field $\mu$SR. In these experiments we follow the angular rotation frequency $\omega$ of the spin of a muon implanted in the sample as a function of temperature. The measurements were performed at the General Purpose Surface-Muon Instrument at the Paul Scherrer Institute (PSI). In Fig. 5(b) we depict the temperature-dependent muon rotation frequency $\omega$ for samples $\alpha$, $\beta$, and $\gamma$. The N\'{e}el temperature is defined as the point where $\omega \rightarrow 0$ upon warming. We find that the N\'{e}el temperatures of the crystals agree with the phase diagram in Fig. 1(a). In particular, as the oxygen composition increases (samples $\alpha$ and $\gamma$), $T_{N}$ decreases. At constant $y$ (samples $\alpha$ and $\beta$) $T_{N}$ increases with $x$. Most importantly, at constant effective doping (samples $\gamma$ and $\beta$) $T_{N}$ increases from 376 K at $x=0.1$ to 420 K at $x=0.4$.

The mechanism for a two-magnon Raman scattering process in an undoped sample corresponds to a simultaneous exchange of two neighboring spins, induced by the incoming photons. As a result, the magnetic coupling of the exchanged spins to their neighboring spins will be broken, and photons with reduced energy will be emitted. Hence, the photon energy shift $E_{max}$ is related to $J$. A simple broken-bond counting argument for spin $1/2$ on a two-dimensional square lattice indicates that the Raman shift peaks at $3J$. A more detailed calculation indicates that the exact factor between the Raman peak and $J$ ranges from $2.71$,~\cite{Weber} via $3.32$,~\cite{Chubukov} to $3.38$.~\cite{Canali} For the purpose of the work presented here it is sufficient to assume that the ratio between the Raman shift and $J$ is close to 3 and constant for all CLBLCO compounds. In a doped sample the number of broken bonds in the scattering process is smaller, and $E_{max}$ is expected to decrease with doping.

\begin{figure}[tbp]
\label{figure6}
\centering
\includegraphics[width=8cm]{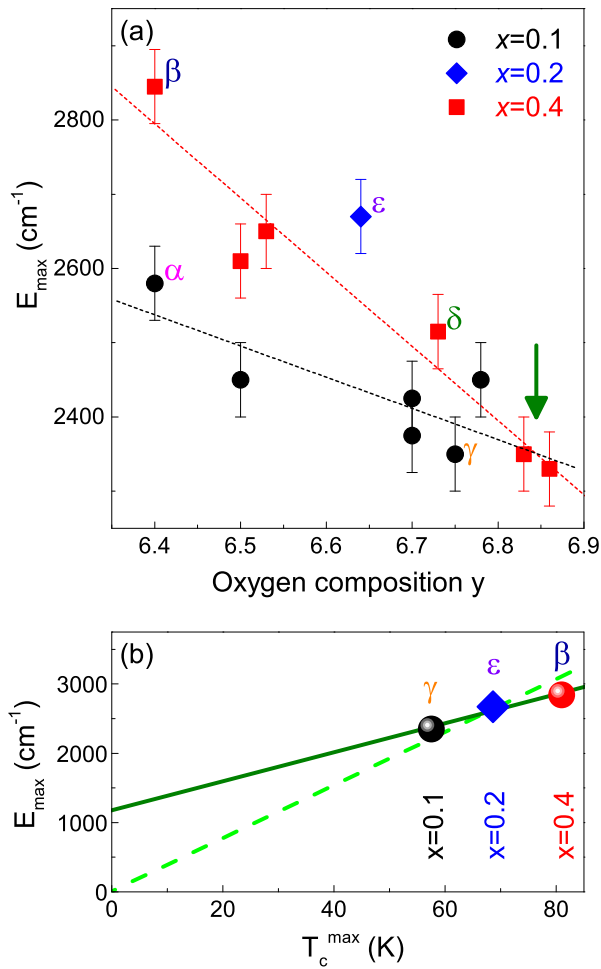}
\caption{(Color online) (a) Doping dependence of the two-magnon mode energy $E_{max}$ for the families $x=0.1$ (black) and $x=0.4$ (red). Samples $\alpha$ -- $\delta$, as well as sample $\varepsilon$ from the $x=0.2$ family, are indicated. The green arrow denotes the crossing regime. (b) $E_{max}$ measured for the $\Delta n_{p\sigma}\simeq 0$ samples $\gamma$, $\varepsilon$, and $\beta$ as a function of $T_{c}^{max}$ obtained from the phase diagram in Fig. 1(a) for the optimally doped samples of the $x=0.1$, $x=0.2$, and $x=0.4$ families. The solid line is a linear fit through the data points; the dashed line denotes a proportionality behavior between $T_{c}^{max}$ and $J$.}
\end{figure}

In Fig. 6(a) we compare the Raman shift $E_{max}$ as a function of oxygen composition $y$ for the two families. The data points corresponding to samples $\alpha$ -- $\delta$ are marked accordingly. Sample $\varepsilon$ of the $x=0.2$ family is also indicated. For low $y$ values, the $x=0.4$ samples have higher two-magnon mode energies than the $x=0.1$ samples. This result indicates that $J$ is larger for the family with higher $T_{c}^{max}$. In both cases a clear and steady decrease is observed as a function of $y$. Around $y=6.8$ there is a crossing point that suggests that $E_{max}$ for $x=0.4$ becomes smaller than that for $x=0.1$. The same trend can be seen for $T_{N}$ in the phase diagram in Fig. 1(a), but the crossing is taking place at a different $y$. Here we note that the two-magnon mode energy $E_{max}$ and $T_{N}$ are not directly related; however, a similar trend is to be expected. The fact that $E_{max}$ shows a different doping dependence for the two families is another indication that the efficiency of doping holes in the CuO$_{2}$ planes is not the same for the two families. The strong doping dependence of $E_{max}$ and the crossing point shown in Fig. 6 emphasize the importance of comparing samples with the same effective doping $\Delta n_{p\sigma}$.

Figure 6(b) summarizes the main finding of this work. In this figure we plot the Raman shift $E_{max}$ obtained at $T=20$ K for the $x=0.1$ family (black), the $x=0.2$ family (blue), and the $x=0.4$ family (red) at $\Delta n_{p\sigma}\approx 0$ versus $T_{c}^{max}$ obtained from the phase diagram in Fig. 1(a). The solid line is a linear fit through the data. A dashed proportionality line through the origin is also shown. Although single crystals from only three families are available for this plot, there is clear evidence for a strong correlation between the magnetic exchange interaction $J\simeq E_{max} /3$, measured by two-magnon Raman scattering, and the superconducting temperature $T_{c}^{max}$, which is close to proportionality.

In a recent Raman study on $R$(Ba,Sr)$_{2}$Cu$_{3}$O$_{y}$ [$R$=(La, ... Lu,Y)], it was found that $T_{c}^{max}$ anticorrelates with $J$ set by internal pressure, and correlates with $J$ induced by external pressure.~\cite{MallettCM12} The internal pressure experiment is in strong contrast to our findings. We believe that the impact of disorder on $T_{c}$ is at the heart of this contradiction. Clarifying this contradiction, which stems from the same experimental method, is essential for understanding the relevant mechanisms for cuprate superconductivity.

\section{Summary}

We report evidence for a strong correlation between the superconducting temperature $T_{c}^{max}$ and the magnetic exchange interaction $J$ in the cuprate system (Ca$_{x}$La$_{1-x}$)(Ba$_{1.75-x}$La$_{0.25+x}$)Cu$_{3}$O$_{y}$, confirmed by two independent, complementary techniques. The two-magnon Raman scattering technique measures $J$ directly but is limited to the antiferromagnetic phase and to single crystals. The analysis of the $\mu$SR technique requires theoretical modeling to extract $J$, but it is applicable from the antiferromagnetic phase through the spin-glass phase and up to the mixed spin-glass superconducting phase.

The exchange interaction $J$ is determined by overlaps of orbitals on neighboring sites. The larger the overlap is, the easier it is for holes to hop from site to site. Thus, in the $x=0.4$ family the overlap of orbitals is larger than in $x=0.1$. This finding is supported by recent angle-resolved photoemission spectroscopy measurements.~\cite{Drachuck13}

\begin{acknowledgments}
This work was supported by the German-Israeli Foundation (GIF, 1171-189.14/2011) and the joint German-Israeli DIP Project (YE 120/1-1 LE 315/25-1). The authors wish to thank the PSI staff for support with the μSR experiments and K. S. Kim for helpful discussions.
\end{acknowledgments}


\begin{thebibliography}{99}

\bibitem{Rinat08}
R. Ofer, A. Keren, O. Chmaissem, and A. Amato, Phys. Rev. B \textbf{78}, 140508(R) (2008).

\bibitem{Kanigel02}
A. Kanigel, A. Keren, Y. Eckstein, A. Knizhnik, J. S. Lord, and A. Amato, Phys. Rev. Lett. \textbf{88}, 137003 (2002).

\bibitem{Rinat06}
R. Ofer, G. Bazalitsky, A. Kanigel, A. Keren, A. Auerbach, J. S. Lord, and A. Amato, Phys. Rev. B \textbf{74}, 220508(R) (2006).

\bibitem{Yaki99}
A. Knizhnik, Y. Direktovich, G. M. Reisner, D. Goldschmidt, C. G. Kuper, and Y. Eckstein, Physica C \textbf{321}, 199 (1999).

\bibitem{Xray}
S. Agrestini, S. Sanna, K. Zheng, R. De Renzi, E. Pusceddu, G. Concas, N. L. Saini, A. Bianconi, Journ. Phys. Chem. Solids \textbf{75}, 259 (2014).

\bibitem{KerenNJP09}
A. Keren, New J. Phys. \textbf{11}, 065006 (2009); T. Cvitani\'{c}, D. Pelc, M. Po\v{z}ek, E. Amit, and A. Keren, Phys. Rev. B \textbf{90}, 054508 (2014).

\bibitem{Eran10}
E. Amit, A. Keren, Phys. Rev. B \textbf{82}, 172509 (2010).

\bibitem{molegraaf}
see, e.g., H. J. A. Molegraaf, C. Presura, D. van der Marel, P. H. Kes, M. Li, Science \textbf{295}, 2239 (2002).

\bibitem{Crystal}
G. Drachuck, M. Shay, G. Bazalitsky, R. Ofer, Z. Salman, A. Amato, C. Niedermayer, D. Wulferding, P. Lemmens, A. Keren, J. Supercond. Nov. Magn. \textbf{25}, 2331 (2012).

\bibitem{Sugai}
S. Sugai, H. Suzuki, Y. Takayanagi, T. Hosokawa, and N. Hayamizu, Phys. Rev. B \textbf{68}, 184504 (2003).

\bibitem{TDepRaman}
M. Bloch, J. Appl. Phys. \textbf{34}, 1151 (1963); S. R. Chinn, R. W. Davies, and H. J. Zeiger, Phys. Rev. B \textbf{4}, 4017 (1971).

\bibitem{Freitas}
P. J. Freitas and R. R. P. Singh, Phys. Rev. B \textbf{62}, 5525 (2000).

\bibitem{Fleury}
P. A. Fleury and R. Loudon, Phys. Rev. \textbf{166}, 514 (1968).

\bibitem{Weber}
W. H. Weber and G. W. Ford, Phys. Rev. B \textbf{40}, 6890 (1989).

\bibitem{Chubukov}
A. V. Chubukov and D. M. Frenkel, Phys. Rev. B \textbf{52}, 9760 (1995).

\bibitem{Canali}
C. M. Canali and S. M. Girvin, Phys. Rev. B \textbf{45}, 7127 (1992).

\bibitem{MallettCM12}
B. P. P. Mallett, T. Wolf, E. Gilioli, F. Licci, G. V. M. Williams, A. B. Kaiser, N. W. Ashcroft, N. Suresh, and J. L. Tallon, Phys. Rev. Lett. \textbf{111}, 237001 (2013).

\bibitem{Drachuck13}
Gil Drachuck, Elia Razzoli, Rinat Ofer, Galina Bazalitsky, R. S. Dhaka, Amit Kanigel, Ming Shi, and Amit Keren, Phys. Rev. B \textbf{89}, 121119(R) (2014).

\end{thebibliography}
\end{document}